\documentclass{aa} 

\usepackage{graphicx}
\usepackage[varg]{txfonts}
\usepackage{longtable,lscape}
\usepackage{natbib}
\bibpunct{(}{)}{;}{a}{}{,}
\usepackage[usenames]{color}

\begin{document}
 \title{CoRoT\,102749568: mode identification in a $\delta$ Scuti star based on regular 
spacings\thanks{The CoRoT space mission was developed and is operated by the French space agency 
CNES, with participation of ESA's RSSD and Science Programmes, Austria, Belgium, Brazil, 
Germany, and Spain. Discovery and multi-colour follow-up observations were obtained at Piszk\'estet\H{o}, 
the mountain station of Konkoly Observatory. Spectra were acquired using the AAOmega 
multi-object spectrograph mounted at the 3.9-m telescope of the Australian Astronomical Observatory in programme 07B/040 and 08B/003 and the HERMES
spectrograph, installed at the 1.2-m Mercator Telescope, operated on
the island of La Palma by the Flemish Community. 
}}

\author{M.~Papar\'o\inst{1}\fnmsep\thanks{\email{paparo@konkoly.hu}}
         \and
              Zs.~Bogn\'ar\inst{1}
         \and
              J.~M.~Benk\H{o}\inst{1}
         \and
              D.~Gandolfi\inst{2} 
         \and
              A.~Moya\inst{3}
         \and
              J.~C.~Su\'arez\inst{4} 
 	\and
              \'A.~S\'odor\inst{1,5}
 	\and
              M.~Hareter\inst{1}
         \and
              E.~Poretti\inst{6}
	\and
              E.~W.~Guenther\inst{7}
 	\and
              M.~Auvergne\inst{8}
	\and
              A.~Baglin\inst{8}
	\and
              W.~W.~Weiss\inst{9}
}

  \offprints{M.~Papar\'o}

  \institute{
    Konkoly Observatory, MTA CSFK, Konkoly Thege M. \'ut 15-17., H-1121 Budapest, Hungary
\and
    Research and Scientific Support Department, ESTEC/ESA, PO Box 299, 2200 AG, Noordwijk, The Netherlands
\and
    Departamento de Astrof\'{\i}sica, Centro de Astrobiolog\'{\i}a (INTA-CSIC), PO Box 78, 28691 
Villanueva de la Ca\~{n}ada, Madrid, Spain
\and
    Instituto de Astrof\'{\i}sica de Andaluc\'{\i}a (CSIC), Glorieta de la Astronom\'{\i}a S/N, 18008, Granada, Spain
\and
    Royal Observatory of Belgium, Ringlaan 3, B-1180 Brussels, Belgium
\and
     INAF - Osservatorio Astronomico di Brera, Via E. Bianchi 46, 23807 Merate (LC), Italy
\and
    Th\"uringer Landessternwarte Tautenburg, Sternwarte 5, 07778, Tautenburg, Germany
\and
     LESIA, Universit\'e Pierre et Marie Curie, Universit\'e Denis Diderot, Observatoire de
     Paris, 92195 Meudon Cedex, France
\and
    Institute of Astronomy, University of Vienna, T\"urkenschanzstrasse 17, A-1180 Vienna, Austria
}

\date{Received; accepted }

\abstract
 { 
The high accuracy of space data increased the number of the
periodicities determined for pulsating variable stars, but the mode
identification is still a critical point in the non-asymptotic regime.
}
 {
We use regularities in frequency spacings for identifying the pulsation
modes of the recently discovered $\delta$ Sct star ID 102749568.
}
 { 
In addition to analysing CoRoT light curves (15 252 datapoints
spanning 131 days), we obtained and analysed both spectroscopic and extended
multi-colour photometric data. We applied standard tools (MUFRAN, Period04,
SigSpec, and FAMIAS) for time-series analysis.
}
{
A satisfactory
light-curve fit was obtaining by means of 52 independent modes and 15
combination terms. The frequency spacing revealed distinct
peaks around large (25.55-31.43 $\mu$Hz), intermediate (9.80, 7.66 $\mu$Hz), and low
(2.35 $\mu$Hz) separations. We directly identified 9 modes,
and the \textit{l} and \textit{n} values of other three modes were extrapolated. The combined
application of spectroscopy, multi-colour photometry,
and modelling yielded the precise physical parameters and confirmed the
observational mode identification. The large separation
constrained the $\log g$ and related quantities. The 
dominant mode is 
the radial first overtone.
}
{}


\authorrunning{M. Papar\'o et al.}
\titlerunning{CoRoT\,102749568: mode identification in a $\delta$ Scuti star based on spacing}
 
\maketitle

%

\section{Introduction}

One of the challenges in stellar astrophysics is to determine the internal
structure of stars. This can be done by studying their pulsation spectrum,
a technique known as asteroseismology. The pulsations of some classes of stars,
such as the Sun and solar-type stars, as well as the white dwarfs, can be
interpreted by using the asymptotic theory of non-radial pulsations
(\citet{Tassoul80}, \citet{Unno89}). Their pulsation modes can then be precisely
identified, leading to the asteroseismological determination of the stellar
structure.

Mode identification is more difficult for $\delta$ Scuti stars because
they are located at the intersection between the classical instability strip
and the main sequence in the HR diagram, a region where the asymptotic theory
of non-radial pulsations is invalid due to low-order modes (complicated by avoided crossing and mixed modes).
Although there are examples of mode identification based on 
long-term ground-based
observations 
(e.g \object{FG\,Vir} -- \citealt{Breger09}, \object{44\,Tau} -- \citealt{Lenz08}), only a few modes have been identified by comparing the observed frequencies with modelling. 

On the theoretical side, \citet{Balona99b} predicted the observation of a larger number of 
high-degree modes and new characteristics once the $\delta$
Scuti stellar light variations are measured at the millimagnitude (mmag) level. 
The space missions, MOST (\citealt{Walker03}, \citealt{Matthews04}), CoRoT \citep{Baglin06}, 
and Kepler (\citealt{Borucki10}, \citealt{Koch10}) resulted accordingly in detecting of a large number of localized peaks in
the Fourier power spectra of the stellar light curves but surprisingly in the
detection of stars with limited frequency content, too,
in both the low-amplitude and high-amplitude $\delta$ Scuti stars, as we now
discuss.

Concerning the low-amplitude $\delta$ Scuti stars (LADS), 
\object{HD\,50844} (\citealt{Poretti09}), \object{HD\,174936} 
(\citealt{Garcia09}), and \object{HD\,50870} (\citealt{Mantegazza12}) 
showed an extremely rich frequency 
content of low-amplitude peaks in the range 0-30~d$^{-1}$. 
A similar dense distribution was obtained in 
\object{KIC\,4840675} (\citealt{Balona12a}). 
However, \object{KIC\,9700322} \citep{Breger11}, one of the coolest $\delta$ Scuti stars 
with $T_{\mathrm{eff}}$ = 6700~K, revealed a remarkably simple frequency content with 
only two radial modes, and a large number of combination frequencies and rotational modulations. 
Based on MOST data, \citet{Monnier10} identified 57 distinct pulsation modes 
in \object{$\alpha$~Oph} above a stochastic granulation noise.

Among the high-amplitude $\delta$ Scuti (HADS) stars, nearly all the light variation of \object{V2367\,Cyg} (\citealt{Balona12b}) is attributed to  
three modes and their combination frequencies. However, 
several hundred other frequencies of very low amplitude are also detected in the star, the effective temperature of which is 
$T_{\mathrm{eff}}$ = 7300~K.  
However, twelve independent terms beside the radial
fundamental mode and its harmonics up to the tenth were identified in the
light of \object{CoRoT\,101155310} (\citealt{Poretti11}). 
Regarding the linear combinations of modes, only 61 frequencies were found down to 0.1~mmag. 
A much
smaller number of low-amplitude modes were thus reported for this HADS star,
although it has the same effective temperature as V2367 Cyg.
As the examples show, the large number of low-amplitude modes were detected by various space missions, and in both of the largest subgroups of $\delta$
Scuti stars.

The question that therefore arises is the following. Are we really detecting
the hundreds of excited modes predicted by theory, or is there a different
cause that would explain the variety of observations? 
Although an 
investigation of the stellar energy balance proved that $\delta$ Scuti 
stars are energetically and mechanically stable even when hundreds of pulsational modes are present 
\citep{Moya10}, the richness can also be 
interpreted as non-radial pulsation superimposed on granulation noise 
\citep{Kallinger10}.

It thus seems that we have reached a level of 
precision where the periodicities stemming from different physical processes
can make the interpretation quite difficult. The appearance of the non-radial modes 
and granulation noise seem to wash out the physical separation of the LADS and HADS groups. 
It may be that only the selection mechanism of the excited modes is different in the two groups. 
However, we do not exactly know the nature of the selection mechanism. 
Any step towards understanding the selection mechanism of non-radial modes 
in $\delta$ Scuti stars would therefore be very valuable. 
A meaningful direction is to find some regularities, if there are any, among the increased number 
of observed frequencies based on a promising parameter such as the frequency spacings.

Some regularities in the spacings of $\delta$ Scuti stars were previously used from both the observational and the theoretical sides. 
Historically, radial fundamental and first and second overtones were 
reported (\citealt{wizinowich79}, \citealt{lopez87} and \citealt{Poretti92}), and an 
additional radial fundamental to first overtone period ratio was found in 44 Tau. 
The latest modelling \citep{Lenz10} has confirmed the existence of the radial fundamental 
and first overtone among the 
observed frequencies, but the previously interpreted second overtone turned out 
to be an $\ell=1$, $p3$ mode. The additional radial sequence is 
identified as $\ell=1$, $p1$, and $p2$ modes. They are interpreted as trapped modes, 
the non-radial counterparts of radial modes \citep{Lenz08}.
The regular sequences can be used for a more accurate characterization of the general 
properties of A-F pulsating stars.  
\citet{Garcia09} interpreted the quasi periodicities in the 
frequencies as a signature of the large separation. 
In the following, we present frequency analyses for the $\delta$ Scuti star \object{CoRoT\,102749568} and 
an extended search for the regularities in the frequency spacings.

\section{Observations and data processing}

The CoRoT target 102749568 was discovered to be variable in the framework of  
CoRoT PECS project. It is identified with the No.~21\footnote{http://www.konkoly.hu/HAG/research/sel21.html} in the field of \object{HD\,292192}. 
We could detect a frequency of 8.70~d$^{-1}$ ($\pm1$~d$^{-1}$ alias uncertainty)
and 
amplitude of 0.02156~mag,  
but no other periodicity was found in the short discovery run.

 \begin{figure}
 \centering
\includegraphics[angle=00,width=7cm]{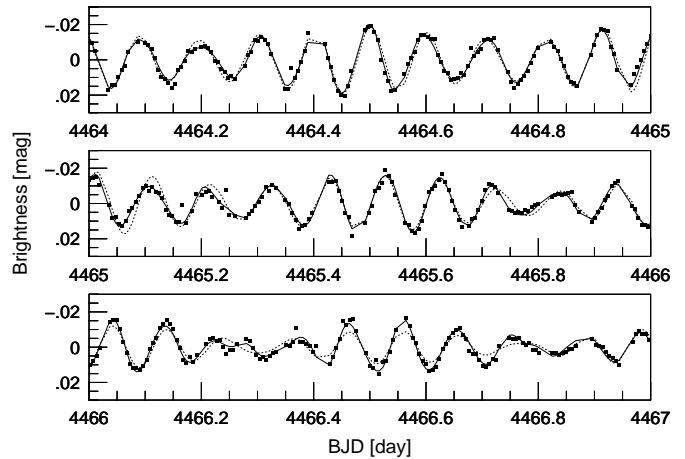}
 \caption{Light curve of CoRoT target 102749568 (in white light) and fit 
with 3 (dashed line) and with 67 frequencies (continuous line). 
}
  \label{fig:rdate}
  \end{figure}

CoRoT\,102749568 ($\alpha = 6^{\mathrm h}44^{\mathrm m}45\fs 98$, 
$\delta = -0\degr 03\arcmin 37\farcs 318$, at epoch: 2000) was 
observed during the first long run in the anti-centre direction (LRa01) from 
October $24^{\mathrm{}}$, 2007 to March $3^{\mathrm{}}$, 2008 which means over 131 days. 
According to the Exo-Dat\footnote{http://cesam.oamp.fr/exodat/}, the CoRoT/Exoplanet input catalogue 
\citep{Deleuil09}, its brightness and colour index were {\it B\,}=15.006 and 
{\it B}$-${\it V}=0.598, and the spectral type and luminosity class are A5 IV (SED).
Additional {\it r} and {\it i} values are given as 14.145 and 13.777 magnitudes.
The Exo-Dat catalogue gives $T_{\mathrm{eff}}$=7091~K for the temperature and {\it E(B$-$V)}=0.45
for the colour excess. The {\it JHK}\/ 2MASS values are 13.128, 12.912, and 12.802 mag. 
Contamination factor is rather low, less than 1 percent. The star is a $\delta$ Scuti type pulsator with a high probability (70\%, \citet{Debosscher09}). 
Other possibilities are an sdBV type (11\%) or an elliptical variable (9\%), but colour indices and short period, respectively, ruled out these hypotheses. 
CoRoT observations were supported by ground-based  
multi-colour follow-up observations and spectroscopy.

\subsection{Raw data}

Corresponding to the generally used eight-minute time sampling on the exofields, 
21\,219 measurements were gathered and presented to us on N2 (science grade) 
level after reduction with the CoRoT pipeline  
\citep{Samadi06, Auvergne09}. We only used the valid zero-flag measurements 
(15\,423, 72.7\%).  
CoRoT\,102749568 
was observed in three CoRoT colours. Some additional 
obvious outliers  were also omitted (15\,252, 71.9\%) from each colour. 
The CoRoT colour channels are described in detail by \citet{Paparo11}. In agreement with the 
CoRoT colour definition, the flux values are the highest ($\approx$70\,000 units) in red, 
about 15\,000 units in green, and about 21\,000 units in blue channels. 
The CoRoT colours were successfully applied to stars 
(CoRoT\,101155310 -- \citealt{Poretti11} and \object{CoRoT\,102781750} -- \citealt{Paparo11}) and 
to exoplanets \citep{Borsa13}. In this paper we compare only the "CoRoT colour indices"
to our multi-colour photometry.
The white light fluxes that were summed up from the three colour channels yield the highest precision.
A moving boxcar method \citep{Chadid10}
was used to get rid of the slight trends and a few jumps in each colour. 
At the same step we converted the fluxes to magnitude. We also transformed the JD times to
Barycentric Julian Dates in Barycentric Dynamical Time ($\mathrm{BJD_{TDB}}$) using 
the applet of \citet{eastman1}. The quality of the white light curve obtained by CoRoT 
is given in Fig.~\ref{fig:rdate}. We explain the least-squares solutions in Sect.~\ref{ssec:freqanal} 

\subsection{Multi-colour photometry from the ground}

\begin{table}
 \caption{Journal of ground-based photometric observations on CoRoT\,102749568 
obtained in the {\it B}\/ band.}
 \label{tabl:log}
 \centering
 \begin{tabular}{rlcrr}
  \hline\hline
Run & UT date & Start time & Points & Length\\
No. & & (BJD-2\,450\,000) & & (h)\\
  \hline
01 & 2011 Feb 05 & 5598.308 & 22 & 2.14\\
02 & 2011 Feb 07 & 5600.249 & 54 & 5.19\\
03 & 2011 Feb 09 & 5602.231 & 44 & 5.55\\
04 & 2011 Nov 24 & 5890.480 & 49 & 3.54\\
05 & 2011 Nov 28 & 5894.458 & 67 & 4.69\\
06 & 2011 Nov 29 & 5895.454 & 54 & 3.64\\
07 & 2011 Nov 30 & 5896.453 & 10 & 0.59\\
08 & 2012 Jan 29 & 5956.267 & 47 & 5.17\\
09 & 2012 Feb 20 & 5978.295 & 36 & 3.34\\
10 & 2012 Feb 21 & 5979.255 & 44 & 4.23\\
11 & 2012 Feb 22 & 5980.250 & 56 & 4.34\\
Total: & & & 483 & 42.42\\
  \hline
 \end{tabular}
\end{table}

We obtained multi-colour photometric data on CoRoT\,102749568 at the Piszk\'estet\H o mountain station of
Konkoly Observatory. Observations were performed on seven nights in winter of 2010-11
and on four nights in winter of 2011-12.
The observations were made with a 
Princeton Instruments VersArray:1300B back-illuminated CCD camera attached to the
1-m RCC telescope and using standard {\it BVRI}\/ filters. 
Table~\ref{tabl:log} contains the journal of observations. The ground-based observations confirmed that the target was not contaminated by any other star in the environment. 
We carried out the photometric reductions by
using standard IRAF\footnote{IRAF is distributed by the National Optical 
Astronomy Observatories, which are operated by the Association of Universities for
Research in Astronomy, Inc., under cooperative agreement with the National Science
Foundation.} packages. After the aperture photometry of the field stars, we 
determined their instrumental colour indices and checked whether the light curves of
the possible comparison stars were free of any instrumental effects or variability.
Finally we selected two stars as comparisons (\object{USNO-B1.0 0899-0101486} and 
\object{USNO-B1.0 0899-0101543}) and used their average light curve for the differential photometry.
Their average instrumental $b-v$ and $v-i$ colour indices both differ from the
variable stars's with $\approx 0.06$~mag.

We transformed the differential light curves to the standard Johnson-Cousins 
system using standard field observations.
Finally we transformed the JD observational times to $\mathrm{BJD_{TDB}}$.
Figure 2 presents the result of one night's observations.

\begin{figure}
 \centering
\includegraphics[width=9cm]{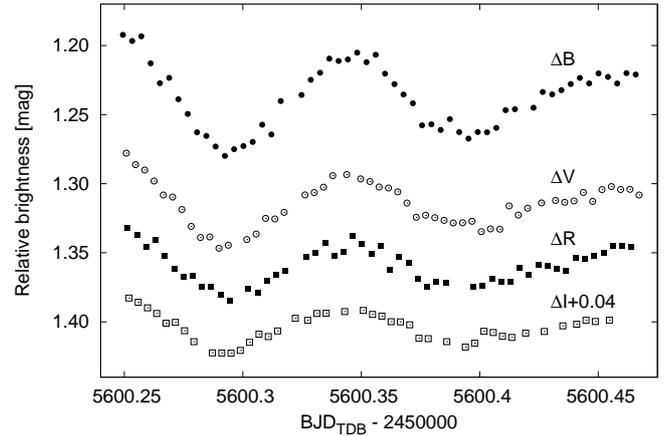}
\caption{{\it BVRI}\/ light curve obtained on the second observing night at Piszk\'estet\H{o}.}
\label{fig:colours}
\end{figure}

We determined the zero-point {\it BVRI}\/ magnitudes of CoRoT\,102749568
by calculating the comparisons' {\it BVRI}\/ values and obtained $B=15.018\pm0.015$,
$V=14.392\pm0.011$, $R=13.919\pm0.011$, and $I=13.510\pm0.019$.
These values are in good agreement with the ones given by the first version of
 Exo-Dat: $B=15.006\pm0.059$ and $V=14.408\pm0.049$.

\section{Pulsational characteristics}\label{sec:pulchar}

\subsection{Frequency analysis of the CoRoT data}\label{ssec:freqanal}

 \begin{figure}
 \centering
\includegraphics[angle=00,width=9cm]{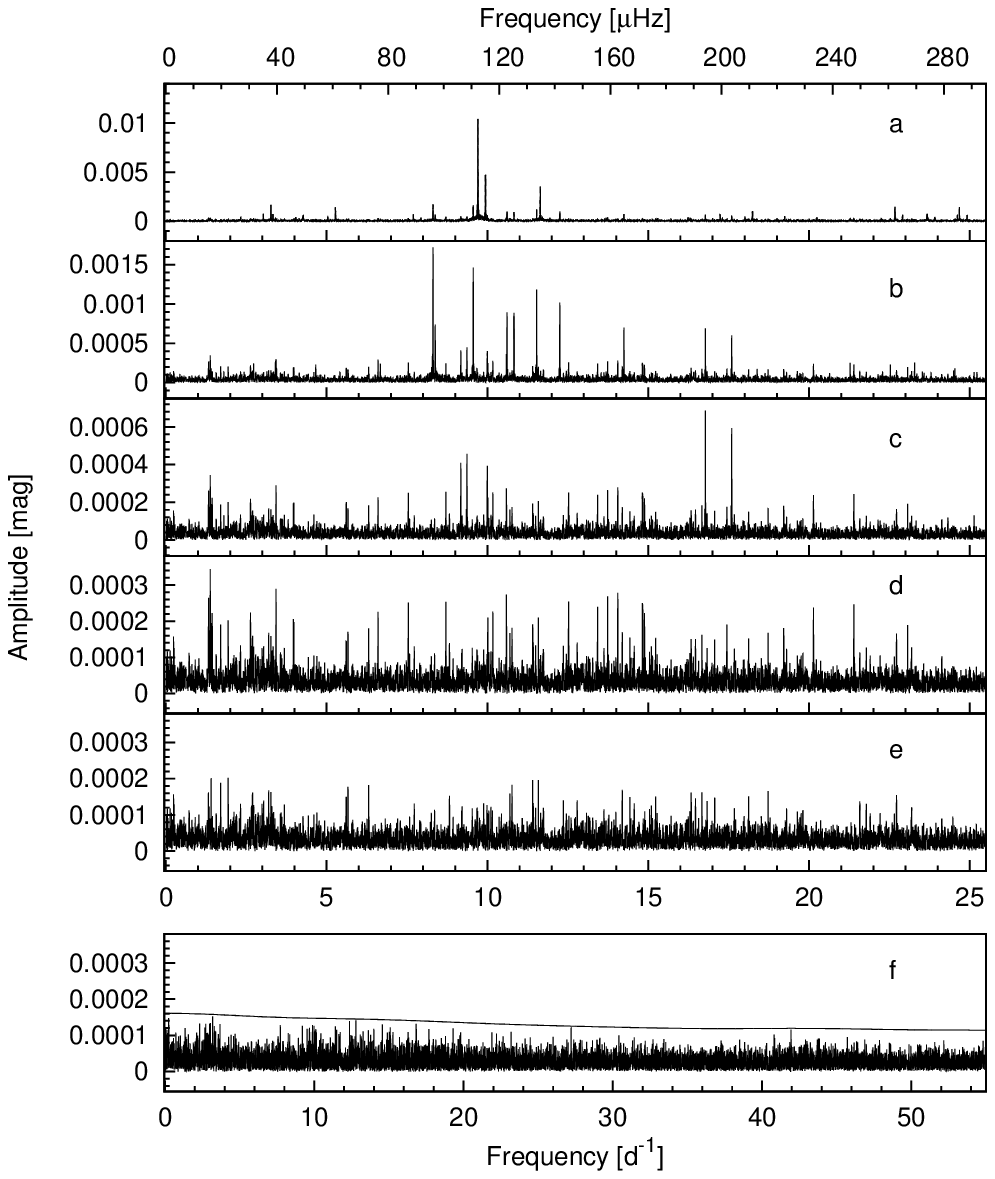}
 \caption{Spectrum at different steps of the frequency search process: 
{\bf a} original spectrum, {\bf b, c, d, e}, and {\bf f} show the spectrum after 
subtracting 3,11,16,41, and 67 peaks, respectively. The residual spectrum is presented until 55 d$^{-1}$ and the significance level is also given over the range. 
 }
  \label{fig:spec}
  \end{figure}

A standard frequency analysis was independently done for each colour by the 
software packages MuFrAn \citep{Kollath90} and SigSpec \citep{Reegen07}. Investigations for amplitude, 
phase variability, significance, and errors were obtained with 
Period04 \citep{Lenz05} and the photometry moduls of FAMIAS \citep{Zima08}. FAMIAS was also used to determine the phase differences and amplitude ratios from the CoRoT and ground-based multi-colour light curve for mode identification. 

\begin{table*}[ht!]
\caption{The 52 independent frequencies of CoRoT\,102749568 to the traditional (S/N=4.0) significance limit. The amplitudes 
and phases are given not only in white light, but also in CoRoT blue, green, and red colours, too. Frequencies with extreme regularities are marked by boldface. The numbering of the frequencies is not continuous, since the combination frequencies are omitted from this table.}
\label{tabl:freq}
\centering
\begin{tabular}{lrcrrrrrrrrrr}
\hline\hline
& \multicolumn{4}{c}{White light} & \multicolumn{2}{c}{Blue} & \multicolumn{2}{c}{Green} & \multicolumn{2}{c}{Red}\\
No. & \multicolumn{1}{c}{Frequency} & \multicolumn{1}{c}{Ampl.} & \multicolumn{1}{c}{Phase} &
\multicolumn{1}{c}{S/N} & \multicolumn{1}{c}{Ampl.} & \multicolumn{1}{c}{Phase} & \multicolumn{1}{c}{Ampl.}
& \multicolumn{1}{c}{Phase} & \multicolumn{1}{c}{Ampl.} & \multicolumn{1}{c}{Phase} \\
& \multicolumn{1}{c}{(d$^{-1}$)} & \multicolumn{1}{c}{(mmag)} & \multicolumn{1}{c}{($1/360\degr$)} & &
\multicolumn{1}{c}{(mmag)} & \multicolumn{1}{c}{($1/360\degr$)} & \multicolumn{1}{c}{(mmag)} &
\multicolumn{1}{c}{($1/360\degr$)} & \multicolumn{1}{c}{(mmag)} & \multicolumn{1}{c}{($1/360\degr$)}\\
\hline
1 &  {\bf 9.70197(1)} & \multicolumn{1}{r}{10.51(3)} & 0.3759(4) & 300.28 & 14.60 & 0.3729 & 12.51 & 0.3764 & 8.80 & 0.3796\\
2 & 9.93909(2) &   4.77 & 0.3616(9) & 136.09 & 6.57 & 0.3467 & 5.60 & 0.3563 & 4.05 & 0.3716\\
3 & {\bf 11.64346(3)} & 3.66 & 0.302(1) & 102.40 & 5.01 & 0.2960 & 4.28 & 0.3064 & 3.09 & 0.3240\\
4 & {\bf 8.30727(6)} &   1.75 & 0.086(2) & 49.98 & 2.43 & 0.0837 & 2.11 & 0.0911 & 1.46 & 0.0880\\
5 & 9.55903(8) &   1.44 & 0.683(3) & 41.54 & 2.01 & 0.6833 & 1.75 & 0.6848 & 1.20 & 0.6836\\
6 & 11.53036(9) & 1.21 & 0.610(3) & 33.86 & 1.60 & 0.6010 & 1.49 & 0.6068 & 1.02 & 0.6143\\
7 & {\bf 12.2485(1)} & 1.01 & 0.411(4) & 27.90 & 1.41 & 0.4125 & 1.19 & 0.4115 & 0.86 & 0.4127\\
8 & {\bf 10.8256(1)} & 0.92 & 0.821(4) & 25.77 & 1.22 & 0.8033 & 1.10 & 0.8199 & 0.78 & 0.8149\\
9 & 10.6072(1) & 0.88 & 0.826(5) & 24.95 & 1.23 & 0.8074 & 1.05 & 0.8050 & 0.73 & 0.8220\\
10 & 8.37540(1) & 0.75 & 0.829(6) & 21.28 & 1.03 & 0.8382 & 0.90 & 0.8134 & 0.63 & 0.8282\\
11 & {\bf 14.2435(2)} & 0.73 & 0.409(6) & 21.02 & 1.04 & 0.3999 & 0.83 & 0.4330 & 0.60 & 0.4284\\
12 & {\bf 16.7771(2)} & 0.68 & 0.032(6) & 19.68 & 0.87 & 0.0367 & 0.76 & 0.0341 & 0.61 & 0.0497\\
13 & {\bf 17.5955(2)} & 0.60 & 0.384(7) & 17.69 & 1.02 & 0.3727 & 0.70 & 0.3614 & 0.45 & 0.3859\\
14 & 9.3633(3) & 0.43 & 0.15(1) & 12.48 & 0.65 & 0.1409 & 0.54 & 0.1589 & 0.35 & 0.1562\\
15 & 9.9970(3) & 0.41 & 0.35(1) & 11.76 & 0.51 & 0.3581 & 0.53 & 0.3210 & 0.35 & 0.3557\\
16 & 9.1715(3) & 0.39 & 0.62(1) & 11.26 & 0.53 & 0.5981 & 0.48 & 0.6081 & 0.34 & 0.6374\\
17 & 14.0508(4) & 0.29 & 0.90(1) & 8.42 &  &  & 0.35 & 0.9325 & 0.29 & 0.8863\\
20 & {\bf 13.7356(4)} & 0.27 & 0.53(2) & 7.73 & 0.31 & 0.5571 & 0.34 & 0.5693 & 0.24 & 0.5229\\
22 & 10.5891(4) & 0.27 & 0.11(2) & 7.56 & 0.30 & 0.0819 & 0.31 & 0.1165 & 0.26 & 0.1303\\
23 & 6.5978(4) &  0.25 & 0.68(2) & 7.40 & 0.45 & 0.6804 & 0.33 & 0.6627 & 0.16 & 0.6827\\
24 & 12.5223(4) & 0.27 & 0.35(2) & 7.39 & 0.35 & 0.3348 & 0.32 & 0.3487 & 0.23 & 0.3542\\
25 & 8.7073(4) &  0.26 & 0.03(2) & 7.29 & 0.39 & 0.0251 & 0.27 & 0.0482 & 0.21 & 0.0372\\
26 & {\bf 7.5406(4)} &  0.25 & 0.96(2) & 7.26 & 0.41 & 0.9493 & 0.25 & 0.9204 & 0.21 & 0.9725\\
27 & 20.1384(5) & 0.24 & 0.25(2) & 7.20 & 0.34 & 0.2581 &  &  & 0.21 & 0.2396\\
28 & 13.4248(5) & 0.23 & 0.52(2) & 6.64 & 0.37 & 0.4827 & 0.34 & 0.5253 & 0.19 & 0.5572\\
29 & 14.8295(5) & 0.23 & 0.77(2) & 6.54 & 0.35 & 0.7904 & 0.25 & 0.7867 & 0.18 & 0.7711\\
30 & 14.8163(5) & 0.23 & 0.32(2) & 6.51 &  &  & 0.26 & 0.2573 & 0.27 & 0.3456\\
31 & 10.1663(5) & 0.23 & 0.33(2) & 6.48 & 0.31 & 0.3343 & 0.28 & 0.3261 & 0.19 & 0.3493\\
33 & 14.8818(5) & 0.22 & 0.55(2) & 6.46 & 0.29 & 0.5412 & 0.22 & 0.5546 & 0.21 & 0.5321\\
34 & 10.0113(6) & 0.22 & 0.67(2) & 6.19 & 0.36 & 0.7020 & 0.26 & 0.6378 & 0.16 & 0.6392\\
38 & 17.4440(6) & 0.20 & 0.07(2) & 5.74 & 0.29 & 0.0638 & 0.24 & 0.0670 & 0.16 & 0.0655\\
40 & 19.2125(6) & 0.18 & 0.34(2) & 5.38 & 0.29 & 0.3964 &  &  & 0.16 & 0.3169\\
42 & 6.3053(6) & 0.18 & 0.30(2) & 5.33 &  &  & 0.23 & 0.3578 & 0.15 & 0.2925\\
43 & 10.7629(6) & 0.19 & 0.37(2) & 5.29 & 0.31 & 0.3681 &  &  & 0.16 & 0.4008\\
45 & 11.4072(6) & 0.19 & 0.63(2) & 5.25 & 0.34 & 0.5835 &  &  & 0.16 & 0.6748\\
46 & 11.5837(6) & 0.19 & 0.38(2) & 5.22 &  &  &  &  & 0.16 & 0.3844\\
47 & 18.7279(7) & 0.17 & 0.93(2) & 5.18 &  &  &  &  & 0.18 & 0.9589\\
49 & 14.1922(6) & 0.17 & 0.95(2) & 4.92 & 0.31 & 0.8875 & 0.21 & 0.9290 & 0.13 & 0.9910\\
50 & 5.6627(6) &  0.18 & 0.33(2) & 4.91 &  &  & 0.24 & 0.2742 & 0.17 & 0.3547\\
51 & 22.7201(7) & 0.16 & 0.58(3) & 4.88 &  &  &  &  & 0.21 & 0.5800\\
53 & 10.6974(7) & 0.17 & 0.74(2) & 4.72 & 0.31 & 0.7431 &  &  &  & \\
54 & 16.6673(7) & 0.16 & 0.71(3) & 4.67 &  &  &  &  & 0.17 & 0.6971\\
55 & 21.5762(7) & 0.15 & 0.13(3) & 4.63 &  &  &  &  & 0.14 & 0.1489\\
56 & 16.3344(7) & 0.16 & 0.72(3) & 4.57 &  &  &  &  & 0.15 & 0.7376\\
57 & 17.0643(7) & 0.15 & 0.56(3) & 4.39 &  &  &  &  & 0.17 & 0.5906\\
58 & 5.6105(7) & 0.16 & 0.22(3) & 4.38 &  &  &  &  & 0.15 & 0.2058\\
59 & 15.2310(7) & 0.15 & 0.80(3) & 4.31 &  &  &  &  & 0.14 & 0.7996\\
61 & 18.1188(8) & 0.14 & 0.29(3) & 4.26 &  &  &  &  & 0.13 & 0.2977\\
62 & 21.7796(8) & 0.14 & 0.72(3) & 4.19 &  &  &  &  & 0.15 & 0.7360\\
65 & 8.8190(8) & 0.14 & 0.38(3) & 4.11 &  &  &  &  & 0.13 & 0.3804\\
66 & 14.4294(7) & 0.14 & 0.07(3) & 4.10 &  &  &  &  &  & \\
67 & 16.4689(8) & 0.14 & 0.88(3) & 4.01 &  &  &  &  &  & \\
\hline
\end{tabular}
\end{table*}

The different software packages resulted in the same solution concerning the
frequencies with higher amplitude than 0.15~mmag. This limit represents the 
traditionally used S/N = 4.0 \citep{Breger93} significance limit. The SigSpec code down 
to the suggested value of the $sig$ parameter (5.0) resulted in 81 peaks in {\it r} ($\sigma=0.018$~mmag), 50 peaks in {\it g}, 52 peaks in {\it b} colours (0.03~mmag), and 94 frequencies in white light (0.017~mmag). Table~\ref{tabl:online} electronically lists the 94 frequencies, 
their amplitude, phase and signal-to-noise ratio values, and the
amplitudes and phases of the corresponding peaks in the CoRoT colours.
The formal errors calculated by Period04
are given in parentheses. 

To achieve the highest probability of finding the regularities between the peaks, 
we present our frequency analyses on the white light results, where the 
low-amplitude, probably non-radial modes were more clearly obtained. The 0.0076 d$^{-1}$ Rayleigh frequency of the whole data set guaranteed that the frequencies were properly resolved. 
The different steps of the period searching process is shown in Fig.~\ref{fig:spec}. 
For better visibility, only the 0 - 25~d$^{-1}$ range is displayed in the first five panels, since there are only orbital peaks 
above the noise level out of this range.
Panel {\bf a} gives the original spectrum dominated by a single high-amplitude 
peak at 9.70197~d$^{-1}$ and two lower amplitude peaks at 9.93909 and 11.64346~d$^{-1}$ values. 
Panels {\bf b}, {\bf c}, {\bf d}, {\bf e}, and {\bf f} show the spectrum after 
subtracting 3, 11, 16, 41, and 67 frequencies, respectively.
The residual spectrum in panel {\bf f} is presented in the 0-55 d$^{-1}$ region and the significance level is also given over the range.
After excluding the possible linear combinations, 52 independent frequencies were found.

Table~\ref{tabl:freq} gives the list of these 52 independent frequencies. The 
amplitudes and phases of the blue, green, and red colours are presented in 
addition to those in white light. 
The fits in Fig.~\ref{fig:rdate} with 3 
and 67 frequencies nicely show that most of the pulsational energy is 
concentrated in the first three modes. The remaining frequencies give only 
minor 
contribution to the light variation but we are able to detect them due to the high-precision measurements. 
The low-amplitude modes facilitate the detection of 
any kind of systematic spacing or other regularities among the frequencies. 

The frequencies are not randomly distributed, as shown in Table~\ref{tabl:group}.
The mode density (8-10 modes/d$^{-1}$) is significantly higher than the typical value 
for $\delta$ Scuti stars (5/d$^{-1}$) given by \citet{Balona11}. These groups contain 32 modes, 
which means 63\% of the independent modes.

\begin{table}
 \caption{Grouping around 6 high-amplitude frequencies.}
 \label{tabl:group}
 \centering
 \begin{tabular}{rrr}
  \hline\hline
Frequency & Range & Number \\
(d$^{-1}$) & (d$^{-1}$)  & \\
  \hline
9.70197  & 0.84 & 7 \\
14.24347 & 0.83 & 7 \\
10.82563 & 0.66 & 6 \\
16.77713 & 0.44 & 4 \\
8.30727  & 0.55 & 4 \\
11.64246 & 0.24 & 4 \\
Total: & & 32\\
  \hline
 \end{tabular}
\end{table}

\begin{table}
\caption{
Possible combination frequencies of CoRoT\,102749568 or {\it g} modes. EDS means near equidistant spacing. Two pairs have near radial period ratios. 
}
 \label{tabl:comb}
\centering
\begin{tabular}{lrcrr}
\hline
\hline
& \multicolumn{4}{c}{White light}\\
No. & \multicolumn{1}{c}{Frequency} & \multicolumn{1}{c}{Ampl.} & \multicolumn{1}{c}{Phase} &
\multicolumn{1}{c}{Remarks}\\
& \multicolumn{1}{c}{(d$^{-1}$)} & \multicolumn{1}{c}{(mmag)} & \multicolumn{1}{c}{($1/360\degr$)} & \\
\hline
18 & 1.3807(3) & 0.33(3) & 0.88(1) & EDS \\
19 & 3.4271(4) &  0.31 & 0.78(1) & $f_{37}/f_{19}$=0.768 \\
21 & 21.3933(4) & 0.25 & 0.19(2) & $f_1+f_3$ \\
32 & 1.3317(6) &  0.26 & 0.66(2) & EDS \\
35 & 23.0662(6) & 0.19 & 0.75(2) & 2x$f_6$ \\
36 & 1.4412(5) &  0.24 & 0.07(2) & EDS \\
37 & 2.6323(5) &  0.22 & 0.08(2) & $f_{37}/f_{19}$=0.768 \\
39 & 3.9661(6) &  0.20 & 0.92(2) & \\ 
41 & 3.9804(6) & 0.20 & 0.85(2) & 3x$f_{64}$ \\ 
44 & 1.9396(6) &  0.21 & 0.97(2) & $f_3-f_1$ \\ 
48 & 1.4097(6) &  0.20 & 0.15(2) & EDS \\ 
52 & 1.7036(6) &  0.19 & 0.13(2) & $f_{64}/f_{52}$=0.777 \\ 
60 & 2.6950(7) & 0.16 & 0.36(3) & $f_7-f_5$ \\
63 & 3.2651(7) & 0.16 & 0.99(2) & $f_3-f_{10}$ \\ 
64 & 1.325(1) & 0.16 & 0.19(3) & $f_1-f_{10}$ \\
\hline
\end{tabular}
\end{table}

Possible linear combinations or {\it g} modes are shown 
in Table~\ref{tabl:comb}. The numbering of frequencies are the missing ones in Table~\ref{tabl:freq} according to the decreasing amplitudes. The sum, differences, and harmonics are the typical linear combinations. Four frequencies show almost equidistant spacing in the 22-38 minute range. 
Although it is near to the spacing of {\it g} modes in \object{CoRoT\,105733033}, a $\delta$ Scuti/$\gamma$ Doradus hybrid star (44.24 minutes, \citealt{Chapellier12}) and near radial period ratios of two pairs of frequencies are also noticed, we are not convinced of the hybrid nature of \object{CoRoT\,102749568}. 

\begin{figure*}
 \centering
\includegraphics[width=17.5cm]{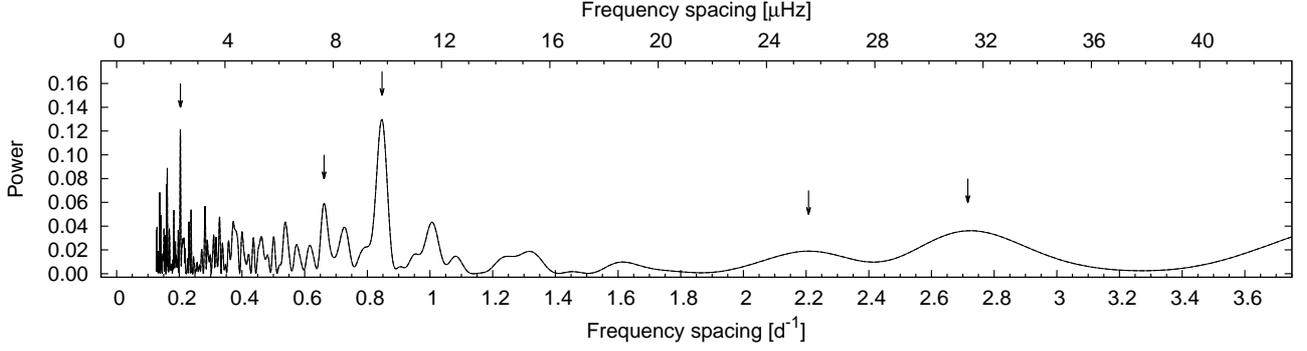}
\caption{Frequency spacing of the 52 independent frequencies determined
in the white light CoRoT data. 
The arrows denote the local maxima at the following
frequency spacing values: 0.2034, 0.6619, 0.8470, 2.2077,
and 2.7158~d$^{-1}$.}
\label{fig:spacings}
\end{figure*}

\begin{table*}
 \caption{ Three sequences are localized according to the largest spacings. 
The zero point of the identification is given by the canonical value of the radial 
fundamental and first overtone in Sequence 1.}
 \label{tabl:seq}
 \centering
 \begin{tabular}{rrrrr|rrrrr|rrrrr}
  \hline\hline
\multicolumn{5}{c}{Sequence 1} & \multicolumn{5}{c}{Sequence 2} & \multicolumn{5}{c}{Sequence 3} \\
\hline
Freq. & Diff. & Ratio & $\ell$ & $n$ & Freq. & Diff. & Ratio & $\ell$ & $n$ & Freq. & Diff. & Ratio & $\ell$ & $n$ \\ 
(d$^{-1}$) & (d$^{-1}$) &  & & & (d$^{-1}$) & (d$^{-1}$) & & & &  (d$^{-1}$) & (d$^{-1}$)  & & &  \\
  \hline
7.54062  &       &       & 0 & 0 & 8.30727  &       &       & 1 & 0 & 11.64346 &       &       & 2 & 0 \\
9.70197  & 2.161 & 0.777 & 0 & 1 & 10.82563 & 2.518 & 0.767 & 1 & 1 & 14.24347 & 2.600 & 0.817 & 2 & 1 \\
12.24848 & 2.547 & 0.792 & 0 & 2 & 13.73559 & 2.910 & 0.788 & 1 & 2 & 16.77713 & 2.534 & 0.849 & 2 & 2 \\
  \hline
 \end{tabular}
\end{table*}

We interpret the regularities of the linear combinations as reflecting the regularities of the {\it p} modes. 
Amplitude variability was checked on 2-3-5-10 day-long subsets. No systematic variation was found for 
the first 20 frequencies.

\subsection{Regular spacing of modes}\label{ssec:spacings}

The search for regular spacings was suggested by pairs of frequencies having very similar ratio to the radial period ratio (0.77).
Following the method used by \citet{handler97} and \citet{Garcia09}, 
we investigated the periodicities in the 52 independent frequencies determined in
 the white light CoRoT data. The input data were the frequencies in the 5.6--22.7 d$^{-1}$ region. The corresponding Nyquist period was 7.92394 d, which means that the spacing was properly resolved above a 0.13 d$^{-1}$ value. The ordinate
unit of the input data of this Fourier analysis is frequency,
therefore the dimension of the Nyquist limit is time.
The resulting FT is shown in Fig.~\ref{fig:spacings}. The 
highest amplitude peaks are at 0.2034, 0.6619, 0.8470, 2.2077, and
2.7158~d$^{-1}$, suggesting very definite spacing amongst the frequencies. We call the attention to the broad double peak at the large value spacing.
In general, the echelle diagram is used in the cases of strictly equidistant spacing (e.g. solar type oscillation). However, in the non-asymptotic regime, the echelle diagram is not the best tool for presenting the regularities\footnote{http://www.iac.es/congreso/cw11/pages/meeting/view-abstract.php?aid=40}, at least for the whole set of frequencies.  
Nevertheless, the systematic spacing leads to regular patterns of the frequencies, for the highest amplitude ones. 
 
 \begin{figure}
 \centering
\includegraphics[angle=00,width=7cm]{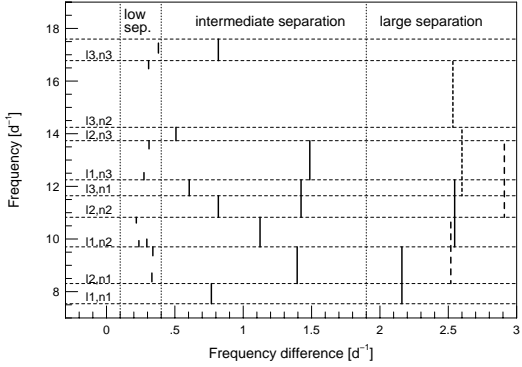}
 \caption{Patterns recognized among the highest amplitude modes are presented. 
The horizontal dashed lines give the frequencies marked by boldface in Table~\ref{tabl:freq}. The frequency differences corresponding to the dominant spacings in Fig.~\ref{fig:spacings} are marked by vertical lines between the given frequencies at the actual values.   
Groups of spacing are divided by vertical dotted lines and labelled as low, intermediate, and large separations. Different sequences in the large separation are plotted by continuous, large dashed, and small dashed lines. The members of sequences are labelled 
by different $\ell$ values and consecutive radial orders ($n$) in the left part.
 }
  \label{fig:group}
  \end{figure}

\begin{table}
 \caption{Mode identification of 12 modes based only on the frequency spacing. 
Italic means the frequencies and their identification extrapolated from the direct determination 
($\ell$=0 and $\ell$=1, $n = 3$ modes and $\ell = 0$, $n = 4$ modes).}
 \label{tabl:reg}
 \centering
 \begin{tabular}{rrrrr}
  \hline\hline
No. & Freq. & $\ell$ & $n$ & Intermediate \\
    &       &        &     &  separation \\
& (d$^{-1}$) &  &  & (d$^{-1}$) \\
  \hline
26 & 7.54062 & 0 & 0 &  \\
4 & 8.30727 & 1 & 0 & 0.767 \\
1 & 9.70197  & 0 & 1 & 1.395 \\
8 & 10.82563 & 1 & 1 & 1.124  \\
3 & 11.64246 & 2 & 0 & 0.817 \\
7 & 12.24848 & 0 & 2 & 0.606 \\
20 & 13.73559 & 1 & 2 & 1.487 \\
11 & 14.24347 & 2 & 1 & 0.508  \\
29 & {\it 14.82949} & {\it 0} & {\it 3} & {\it 0.586} \\
56 & {\it 16.33444} & {\it 1} & {\it 3} & {\it 1.505} \\
12 & 16.77713 & 2 & 2 & 0.443 \\
13 & 17.59551 & {\it 0} & {\it 4} & 0.818 \\
  \hline
 \end{tabular}
\end{table}

Pattern recognition is a well-known technique that was successfully applied to white 
dwarf stars (\citealt{Winget91}, \citealt{Winget94}, \citealt{handler98}). We present in Fig.~\ref{fig:group} the frequency versus frequency difference diagram as a new tool for representing the patterns when the frequencies do not show strictly equidistant spacing.
Along with the central peaks of the main groups (Table~\ref{tabl:group}), dominant peaks of 
some less dense groups were also included (marked by boldface in Table~\ref{tabl:freq}). The structure of Fig.~\ref{fig:group} is described in the caption.
The diversity of the frequency differences clearly show the non-equidistancy that is most remarkable in the large separation region. Nevertheless, it gives very important details.

We can distinguish three sequences of modes 
connected to each other by the large separation. The different sequences cross each other, and are interweaved.  
The highest frequency peak at $f_{13}$=17.595~d$^{-1}$ is not included in any of these sequences. From a theoretical point of view,
the expected large separations of $\delta$ Scuti stars in their last third of
time in the MS range from 2.15~d$^{-1}$ to higher values. Accepting the large frequency differences as the large separation between the consecutive radial orders, we distinguished three sequences containing three consecutive radial orders in each sequence. Three sequences mean that modes with three different $\ell$ values are excited. The labelling of the frequencies in the left part of Fig.~\ref{fig:group} represents this conclusion. 
Since the frequency pairs in Sequence 1 (Table~\ref{tabl:seq}) have radial period ratios that are acceptable for the radial 
overtones, the zero point of the mode identification is fixed at the first label: l1,n1 = 0,0. This means that we identify the frequency at 7.54062 d$^{-1}$ as the radial fundamental mode. The members of sequences, frequency differences between the members, period ratios, and the mode identification are summarized in Table~\ref{tabl:seq}.  

\begin{figure*}
 \centering
\includegraphics[width=16cm]{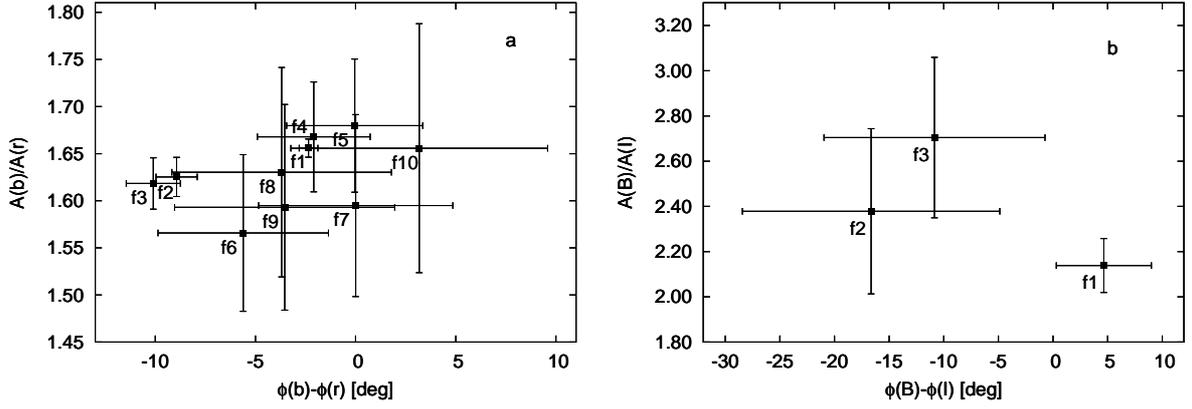}
\caption{Phase differences and amplitude ratios obtained for the largest 
amplitude frequencies in different band passes, {\it b}\/ and {\it r}\/ CoRoT colours, and Johnson--Cousins 
{\it BI}\/.}
\label{fig:phasediff}
\end{figure*}

\begin{table*}
 \caption{Amplitude, phase, and S/N values of the CoRoT dataset's three largest
amplitude modes derived by the multi-colour, ground-based data.}
 \label{tabl:bvri}
 \centering
 \begin{tabular}{rrrrrrrrrrrrr}
  \hline\hline
\multicolumn{1}{c}{Frequency} & \multicolumn{1}{c}{$A_B$} & \multicolumn{1}{c}{$A_V$} & \multicolumn{1}{c}{$A_R$} & \multicolumn{1}{c}{$A_I$} & \multicolumn{1}{c}{$\phi_B$} & \multicolumn{1}{c}{$\phi_V$} & 
\multicolumn{1}{c}{$\phi_R$} & \multicolumn{1}{c}{$\phi_I$} & \multicolumn{1}{c}{S/N$_B$} & \multicolumn{1}{c}{S/N$_V$} & \multicolumn{1}{c}{S/N$_R$} & \multicolumn{1}{c}{S/N$_I$}\\
\multicolumn{1}{c}{(d$^{-1}$)} & \multicolumn{4}{l}{(mmag)} & \multicolumn{4}{l}{($1/360\degr$)} & & \\
\hline
9.70197 & 20.9(7) & 14.8(5) & 11.9(4) & 9.8(4) & 0.221(5) & 0.229(5) & 0.222(6) & 0.208(7) & 13.8 & 12.5 & 13.9 & 11.7\\
9.93909 & 7.8(6) & 6.3(5) & 5.3(4) & 3.3(4) & 0.56(1) & 0.57(1) & 0.58(1) & 0.60(2) & 5.2 & 5.3 & 6.1 & 4.0\\
11.64347 & 10.2(6) & 6.1(5) & 5.5(4) & 3.8(4) & 0.45(1) & 0.45(1) & 0.47(1) & 0.48(2) & 7.0 & 5.5 & 6.4 & 4.9\\
\hline
 \end{tabular}
\end{table*}

Following the general labelling we can state that the intermediate spacings appear between the frequencies with different $\ell$ values, either between the consecutive or the same radial orders. In the non-asymptotic regime we do not see the comb-like distribution of modes with odd and even $\ell$ values. However, we can find some regularities in the intermediate value spacing, too.
In Table~\ref{tabl:reg} the frequencies are listed in the order of increasing values. The identification based on the large separation is also given, together with the intermediate separation of the consecutive frequencies. The frequency difference is given in the line of the higher frequency value.

Larger and smaller intermediate separations regularly follow each other, although the values are not the same. The higher values (such as 1.124, 1.487, and 1.505 d$^{-1}$) appear between the same radial orders of the $\ell = 1$ and $\ell = 0$ modes. The higher values are followed by lower values (0.817, 0.508, and 0.443 d$^{-1}$) of the consecutive radial orders of $\ell = 2$ and $\ell = 1$ modes. The next set of lower values (0.606, 0.586, and 0.818 d$^{-1}$) belong to the second consecutive radial orders of the $\ell = 0$ and $\ell = 2$ modes. 
Comparing the direct identification of nine frequencies based on the large separation (Table~\ref{tabl:seq}), we presented here the identification of 12 modes. 
The regular sequence of the larger and smaller intermediate separations allowed us to identify two modes in the large gap between 14.243 and 16.777 d$^{-1}$ and the highest value frequency in Fig.~\ref{fig:group} that was not included in any of the sequences. These identifications are marked by italics in Table~\ref{tabl:reg}. 
The large separations of these modes to the consecutive radial orders (2.5 or 2.84~d$^{-1}$ for $\ell = 0$ modes and 2.41~d$^{-1}$ for the $\ell = 1$ modes) falls within the large-separation range.

The observational explanation of the low-value spacing is not obvious.
In 
Fig.~\ref{fig:group} we present only those spacings that are around the central frequencies as doublets or triplets. 
A very pronounced example is a double peak at 10.82563 and 10.60715~d$^{-1}$ with the same amplitudes. 
The triplet arrangement of the low-value spacings around the 9.70197~d$^{-1}$ dominant mode that we identified as the radial 
first overtone questions the interpretation as a rotational splitting.

The most remarkable result of our frequency analyses is the identification of twelve modes 
using nothing else than the frequency values and our general knowledge of the pulsation.

\subsection{Multi-colour mode identification}

It is well-known that phase differences and amplitude ratios obtained in different
bandpasses are correlated with the $\ell$ values (see e.g. \citealt{watson88}).
Although CoRoT colours are not calibrated, we tried to use the amplitude and phase differences 
of the modes between {\it b} and {\it r} colours for the mode identification.
 Figure 6a shows the first ten largest amplitude modes in the 
($\phi_b - \phi_r$, $A_b / A_r$)  plane. All of them, except
 $f_{\mathrm{10}}$, have negative phase differences, though $f_{\mathrm{5}}$ and $
f_{\mathrm{7}}$ are close to zero. A positive value implies an $l=0$ mode, but
taking the error bars into account, we cannot state unambiguously that we see
radial mode(s), even in the case of $f_{\mathrm{10}}$. Additionally, $f_{10}$ is only a member 
of the group centred on the higher amplitude peak at 8.307~d$^{-1}$. 

The three largest amplitude modes dominating the CoRoT dataset can be found by
our {\it BVRI} data as well. Table~\ref{tabl:bvri} summarizes their amplitude, phase,
and S/N values. As the follow-up observations show, the 9.702~d$^{-1}$ frequency 
remained the dominant one in the 2011/12 observing season. The amplitude of 11.643~d$^{-1}$ 
frequency became larger than the second highest amplitude mode in CoRoT observation.

All colours (\citealt{Balona99a}, \citealt{Garrido00}) and the {\it B$-$V}\/ colour 
index were used to place the three 
modes on the phase difference versus amplitude ratio plane. Unfortunately, the location of 
modes are different on the different planes, and the error bars are rather large, 
especially for the colour index. The phase difference of the dominant mode 
$f_1$ gives a low negative value both in the {\it B} vs. {\it V}\/ and in the {\it B}\/ vs. {\it R}\/ 
planes. 
In the {\it B} vs. {\it V}\/ plane, $f_2$ is close to it, 
but in the {\it B} vs. {\it R}\/ plane, $f_3$ is close to $f_1$. 
The most definite arrangement of modes was obtained on the ($\phi_B - \phi_I$,
 $A_B / A_I$) plane (Fig.~\ref{fig:phasediff}b). A positive phase difference of $f_1$ is obtained, and even 
the error bar is in the positive region, and $f_2$ and $f_3$ are well-separated in the negative region. 
It is suggested that $f_1$ is a radial mode ($\ell = 0$), while $f_2$ and $f_3$ have different $\ell$ values. 
 
\section{Stellar parameters}

\subsection{Spectroscopy}\label{sect:spec}

A low-resolution spectrum R=1\,300 of CoRoT\,102749568 was acquired in January 
2009 using the AAOmega multi-fibre spectrograph mounted at the 3.9-m telescope 
of the Australian Astronomical Observatory. We refer the reader to \citet{sebastian12} 
and \citet{guenther12} for details on the instrument set-up, observing strategy, 
and data reduction. By comparing the AAOmega spectrum with a grid of suitable templates, 
\citet{guenther12} classified CoRoT\,102749568 as an F1\,IV star. A visual inspection 
of the AAOmega spectrum revealed the presence of a moderate H$\alpha$ emission 
(FWHM=450 m\AA), whose nature will be discussed later. 

Two high-resolution spectra of CoRoT\,102749568 were obtained on two 
consecutive nights in January 2013 using the High Efficiency and Resolution 
Mercator Echelle Spectrograph (HERMES; \citealt{Raskin11}) 
mounted at the 1.2-m Mercator telescope of Roque de los Muchachos Observatory 
(La Palma, Spain). We used the $2.5\,\arcsec$ fibre, which provides a resolving 
power of R=85\,000 and a wavelength coverage of about 3800--9000~\AA. The exposure 
time was set to 1800 and 3600 s on the first and second nights, respectively. The data were reduced using 
the automatic data-processing pipeline of the instrument, which includes bias 
subtraction, flat fielding, order tracing and extraction, wavelength calibration, 
and cosmic-ray removal. The two epoch spectra were corrected for barycentric motion 
and combined into a single co-added spectrum having a S/N ratio 
of about 10, 15, and 20 per pixel at 5000, 6000, and 8500~\AA, respectively. The 
HERMES data revealed a rapidly rotating star with a projected rotational velocity 
of $v$\,sin\,$i_{\star}=115\pm20$~km/s. No radial velocity variation -- at a level of $\sim 4$~km\,s$^{-1}$ -- was found
between the two epoch spectra, which suggests that CoRoT\,102749568 is most 
likely not 
a short-period binary system.
The low S/N of the co-added spectrum, along with the
relatively high $v$\,sin\,$i_{\star}$ did not allow us to perform a meaningful 
spectral analysis. Nevertheless, the high resolution of the HERMES data allowed us to investigate 
the nature of the H$\alpha$ emission better. We found that the latter is relatively narrow (FWHM=0.60~\AA) 
and blue shifted by $\sim$0.5~\AA\ with respect to the core of the H$\alpha$ absorption 
line, suggesting a non-stellar origin.

We used the calibration from \citet{straizys81} to convert the F1\,IV spectral type 
of CoRoT\,102749568 into effective temperature ($T_{\mathrm eff}$) and surface 
gravity ($\log g$). Error bars were estimated taking the uncertainty 
in the spectral classification reported in \citet{guenther12} into account. We then refined the
effective temperature and surface gravity by fitting the AAOmega spectrum with a grid 
of Kurucz stellar models \citep{Kurucz79}, constraining the $T_{\mathrm eff}$ -- $\log g$ parameter 
space around the previously derived values. Assuming a solar metallicity, we obtained
$T_{\mathrm eff}=7000\pm200$~K and $\log g=3.75\pm0.25$.

\begin{figure}
 \centering
\includegraphics[width=9cm]{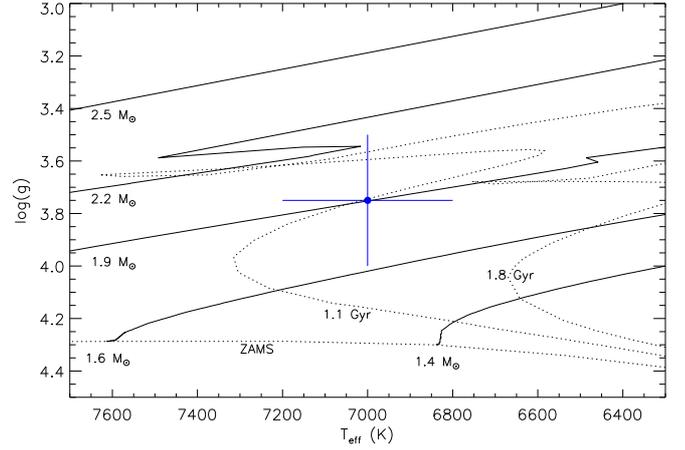}
\caption{Surface gravity vs. effective temperature diagram of
CoRoT\,102749568 (blue dot with error bars). The \citet{girardi00}
evolutionary tracks for 1.4, 1.6, 1.9, 2.2, and 2.5~M$_\odot$ are
over-plotted with continuous lines. Both the zero-age main sequence line,
and the isochrones for 1.1 and 1.8~Gyr are shown with dotted lines.}
\label{fig:logg_vs_teff}
\end{figure}

To estimate stellar mass, radius, luminosity, and age, we compared the location of the 
star on a $\log g$ vs. $T_{\rm eff}$ H-R diagram with the Padova evolutionary tracks 
and isochrones from \citet{girardi00}, as shown in Fig.~\ref{fig:logg_vs_teff}. We
obtained a stellar mass of $M_\star = 1.90\pm0.35$~M$_\odot$ and a stellar radius of 
$3.0^{+1.4}_{-1.0}$~R$_\odot$, with a luminosity of $L=19.5^{+21.0}_{-11.5}$~L$_\odot$ 
and an age of $1.1^{+0.6}_{-0.4}$~Gyr.

\begin{figure}
 \centering
\includegraphics[width=9.3cm]{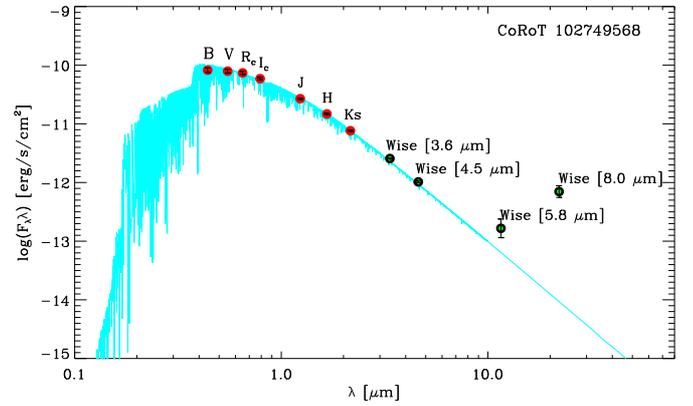}
\caption{Dereddened SED of CoRoT\,102749568. The best-matching NextGen
synthetic spectrum is overplotted with a light blue line. The 
$W_3$ (5.8 $\mu$m) and $W_4$ (8 $\mu$m) WISE magnitudes were not used to
estimate reddening and distance.}
\label{fig:SED}
\end{figure}

\begin{table*}
 \caption{Physical parameters of CoRoT\,102749568 determined
from spectroscopic and photometric observations and modelling.}
 \label{tabl:phparam}
 \centering
 \begin{tabular}{*3{lc}}
  \hline\hline
Parameter\tablefootmark{a} & Value & Parameter\tablefootmark{d} & Value & Parameter\tablefootmark{e} & Value \\
\hline
$T_\mathrm{eff}$\tablefootmark{a} & $7000\pm200$~K & $M_V$ & $1.373\pm0.1$ & $T_\mathrm{eff}$\tablefootmark{e} & $7000\pm2000$~K \\
$\log g$\tablefootmark{a} & $3.75\pm0.25$~dex & $M_{\mathrm{bol}}$ & $1.403\pm0.1$ & $\log g$\tablefootmark{e} & $3.7\pm0.1$ \\
$M_\star$\tablefootmark{b} & $1.90\pm0.35$~M$_\odot$ & $M_\star$ & $2.033\pm1.1$~M$_\odot$ & $M_\star$\tablefootmark{e} & $1.8\pm0.2$~$M_\odot$ \\
$R_\star$\tablefootmark{b} & $3.0^{+1.4}_{-1.0}$~R$_\odot$ & $R_\star$ & $3.156\pm0.2$~R$_\odot$ & 
$R_\star$\tablefootmark{e} & $3.2\pm0.4$~R$_\odot$ \\
$L_\star$\tablefootmark{b} & $19.5^{+21.0}_{-11.5}$~L$_\odot$ & $L_\star$ & $21.454\pm1.8$~L$_\odot$ & 
$L_\star$\tablefootmark{e} & $23\pm8$~L$_\odot$ \\
Age\tablefootmark{b} & $1.1^{+0.6}_{-0.4}$~Gyr &  & &  Age\tablefootmark{e} & $1.7\pm0.6$~Gyr \\
$d$\tablefootmark{c} & $2600^{+1200}_{-900}$~pc &  &  & \\
$A_{\rm v}$\tablefootmark{c} & $0.8\pm0.1$~mag &  & & \\
\hline
 \end{tabular}
\tablefoot{
\tablefoottext{a}{Values from the AAOmega spectrum,}
\tablefoottext{b}{from evolutionary tracks and isochrones,}
\tablefoottext{c}{determined by photometric data and a NextGen synthetic spectrum.}
\tablefoottext{d}{Parameters calculated using $T_\mathrm{eff}$, $\log g$,
and $f_\mathrm{26}$ as a radial fundamental mode.}
\tablefoottext{e}{Range of parameters from modelling}
}
\end{table*}

The interstellar extinction ($A_{\mathrm v}$) and distance ($d$) to CoRoT\,102749568 were 
derived following the method described in \citet{gandolfi08}. We merged the Konkoly optical 
photometry with the infrared 2MASS and Wise data and built up the spectral energy distribution 
(SED) of the target star. The interstellar extinction was then derived by fitting simultaneously 
all the colours encompassed by the SED with synthetic magnitudes, as obtained from a NextGen 
model spectrum \citep{hauschildt99} having the same photospheric parameters as the target star. 
We excluded the $W_3$ (5.8 $\mu$m) and $W_4$ (8 $\mu$m) WISE magnitudes as they are only upper limits. 
Assuming a black body emission at the star's effective temperature and radius, we found 
$A_{\mathrm v}=0.8\pm0.1$\,mag and $d=2.6^{+1.2}_{-0.9}$~Kpc. The dereddened SED of 
CoRoT\,102749568 is shown in Fig.~\ref{fig:SED} along with the NextGen synthetic spectrum. The 
Wise data points at 5.8~$\mu$m and 8.0~$\mu$m show an apparent infrared excess with respect to 
the expected photospheric emission. We believe that this excess, along with the H$\alpha$ emission 
detected both in the AAOmega and HERMES data, is caused by the strong background contamination 
arising from the nearby H\,II region \object{Sh\,2-284}, on which CoRoT\,102749568 is spatially 
projected.

Nevertheless, we checked the possibility that our target belongs to Sh\,2-284, whose heliocentric 
distance is stated in the literature between 4 Kpc \citep{Cusano11} and 6 Kpc \citep{Turbide93}. While 
the latter clearly rules out a cluster membership, the former is consistent within $\sim$1$\sigma$ 
with our determination of the heliocentric distance of CoRoT\,102749568 ($d=2.6^{+1.2}_{-0.9}$~Kpc). 
However, the F-type stars listed by \citet{Cusano11} are three to four magnitudes fainter than our target 
star. This suggests that CoRoT\,102749568 is most likely a foreground object with respect to Sh\,2-284.

\subsection{Multi-colour photometry}

Empirical relations are widely used to constrain some physical parameters. 
The period-luminosity relation for $\delta$ Scuti stars connects the pulsation 
to the stellar physical parameters. We can use it in two way. Knowing the  period 
of the radial fundamental mode, we can get the physical parameters, however, if we 
know the physical parameters from other source (i.e. spectroscopy), then we can use 
the relation for finding the period of the radial fundamental mode and confirm the 
tentative mode identification suggested by multi-colour photometry. Knowing the physical 
parameters from spectroscopy, we followed the second way.

According to Figs.~\ref{fig:phasediff}b and 6a 
the $f_1$, $f_{10}$, {\bf $f_4$}, $f_3$, $f_5$, and $f_7$ frequencies could be radial modes. 
We inserted the period values in the period-luminosity relation of \citet{Poretti08}, 
\[
 M_V = -1.83(\pm0.08)-3.65(\pm0.07)\log P_\mathrm{F},
\]
and $M_V$ = 1.772, 1.539, 1.526, 2.061, 1.749, and 2.141~mag were obtained, respectively. 
Considering these
values and the atmospheric parameters determined by spectroscopy, CoRoT\,102749568 can be found 
both in the $T_\mathrm{eff}$--$M_V$ and $T_\mathrm{eff}$--$\log g$ plane near the 
red edge of the observed instability strip (see e.g. \citealt{McNamara00}, 
\citealt{Uytterhoeven11}) and represents an evolved star from the zero-age main sequence. 
CoRoT\,102749568 is obviously outside of the typical location of the HADS stars.

The luminosity, radius, and mass of CoRoT\,102749568 
were derived using a bolometric correction of $\mathrm{BC}_V = 0.031$ \citep{flower96, torres10}. We got $R_\star$ = 2.626, 2.923, 2.940, 2.298, 2.654, and 2.215~R$_\odot$, 
$M_\star$ = 1.407, 1.744, 1.765, 1.078, 1.438, and 1.001~M$_\odot$, and $L_\star$ = 14.849, 
18.405, 18.626, 11.377, 15.175, and 10.566~L$_\odot$ values. 

Comparing these values to the luminosity, radius, and mass obtained from spectroscopy, 
we conclude that the $f_3$=11.643 and $f_7$=12.248~d$^{-1}$ frequencies are 
definitely excluded as radial fundamental modes. The $f_{10}$=8.375~d$^{-1}$ and $f_4$=8.307~d$^{-1}$ 
frequencies give the closest values to the spectroscopic ones, but the 
dominant mode's ($f_1$=9.702~d$^{-1}$) values are practically in the error bar, 
which means the dominant mode can be a radial mode. The closest values obtained 
from $f_{10}$ and $f_4$ suggest that the dominant mode maybe a radial first overtone, 
supporting the $f_{26}$=7.5406~d$^{-1}$ frequency as radial fundamental mode. 
The stellar parameters calculated from $f_{26}$ as a radial mode are given in 
Table~\ref{tabl:phparam}. The physical parameters agree with the spectroscopic value. 
Owing to the low amplitude of $f_{26}$, the error bars of the amplitude ratio and phase 
differences are so large that $f_{26}$ is not plotted in Fig.~\ref{fig:phasediff}a.
Based on the empirical period-luminosity relation and our spectroscopic investigation, 
we conclude that the dominant mode of CoRoT\,102749568 is the radial first overtone and 
the 7.5406, 9.702, 12.248~d$^{-1}$ sequence contains the radial fundamental and the 
second overtone modes.
Table~\ref{tabl:phparam} summarizes the physical parameters determined
from spectroscopy, photometry, and modelling of CoRoT\,102749568.

\section{Modelling}\label{sec:modelling}

A detailed
modelling of this star is beyond the scope of this work, but we have developed an
analysis that compares different models, and some very interesting 
conclusions can be subtracted from a first study.
We computed 
non-rotating equilibrium and adiabatic oscillation models. We used the evolutionary 
code CESAM \citep{Morel97, Morel08} and the pulsation code GraCo 
\citep{Moya04, Moya08} to calculate the equilibrium models and the adiabatic
frequencies only with $\ell$ = [0, 3], respectively, for visibility reasons 
\citep{dziembowski77}. 
Low-order modes 
must be enough to describe the highest amplitude modes. We constructed a grid of models covering the 
instability strip of the $\delta$ Scuti stars. We computed the range [1.25, 2.20]~M$_\odot$ in mass 
with a step of 0.01 and the range [-0.52, 0.08]~dex in [Fe/H] with a step of 0.2.

We then selected the only models fulfilling the spectroscopic $T_{\mathrm{eff}}$ and
$\log g$ displayed in Table~\ref{tabl:phparam} and additionally let the metallicity of
the models move in the range [-0.1,0.1]~dex. This additional degree of freedom has a 
low impact on the determination of the physical characteristics of the models 
fulfilling observations. 
Our main goal was to check the spacing between 2-3~d$^{-1}$ given in Fig.~\ref{fig:spacings}.
According to \citet{Garcia13} this can be the signature of the large separation.
Taking into account that the rotation velocity deduced in Sect.~\ref{sect:spec} has a
frequency also in this region, we cannot decide that the two peaks between 2-3~d$^{-1}$ are the consequence of the non-equidistant value of the large separation or one of them is a consequence of the rotation.
The width of the peaks is an additional source of
uncertainty. Therefore,
we decided to be conservative and 
estimated that the large separations can be in the range [2.1, 2.75]~d$^{-1}$ for the modelling, but it could be  
half of or double this value.

After modelling we find that there are no models with the observed T$_{\mathrm{eff}}$, $\log g$, 
and near solar metallicity with large separations in the ranges [1.05, 1.37]~d$^{-1}$ and [4.2, 5.5]~d$^{-1}$. 
Therefore, the peaks in Fig.~\ref{fig:spacings} can be a direct measurement of the large 
separations. All models fulfilling the observed photometric uncertainty box, solar metallicity, and 
large separations in the range [2.1, 2.75]~d$^{-1}$ have physical parameters listed in 
Table~\ref{tabl:phparam}. Here we can see that this new observable mainly constrains $\log g$ and 
related quantities.

Another possible source of information is the peak around 0.2~d$^{-1}$ in
Fig.~\ref{fig:spacings}.
As the star evolves, its internal structure changes, and it affects 
the pulsational modes. Some effects appear such as the avoided
crossing. This avoided crossing {\it locally} destroys any regular pattern at
any given spherical degree $\ell$. Therefore, as the average large separation
is similar for every $\ell$, the avoided crossing does not have a strong
impact here, but other useful relations like the small separations are
highly affected by this effect. 
Therefore, we do not expect an easy identification 
of the effect of avoided crossing and the identification of
this spacing as a signature of the small separations is unlikely.
Nevertheless, a comprehensive modelling must be done in order to finally
discard this possibility or to claim that this is the first observational probe of
the small seprations in a $\delta$ Scuti star.

This comprehensive modelling requires including of the effects of
rotation. Assuming the physical parameters
extracted with the FT method, the minimum possible value for the rotational
velocity at equator would imply
a deformation of the star that makes it necessary to use non-peturbative
techniques to calculate the oscillations.
Although the current  non-pertubative methods predict no significant effects
on large spacings, they predict
a redistribution of the modes that is slightly different to pertubative predictions
\citep{Lignieres06, Reese06}.
This combination of both methods  (FT with non-rotating models, and
non-perturbative oscillations based
on 2D models) should be able to provide more insight into the mode
identification for the present star.

\section{Conclusions}

The CoRoT $\delta$ Scuti star 102749568 shows only a limited number of peaks. 
The 52 independent modes are in the region of the radial and non-radial modes typical 
of $\delta$ Scuti stars. Both the observation and the non-rotating equilibrium and adiabatic oscillation models show the grouping of 
the frequencies and allow the determination of the large separation [2.1 - 2.75]~d$^{-1}$ 
in a $\delta$ Scuti star in the non-asymptotic regime. 

The excited modes reveal an unexpected regular arrangement that allowed us to identify 
12 modes based only on the observed frequencies. Five consecutive radial orders of the radial 
modes (from $n = 0$ to $n = 4$) are excited besides the four consecutive radial orders 
of $\ell = 1$ and three consecutive radial orders of $\ell = 2$ non-radial modes. 
The identification of some modes were extrapolated on the basis of the extreme regularity. 
Multi-colour photometry confirmed the identification of the highest amplitude modes. 
Physical parameters were derived from spectroscopy that allowed the inverse application 
of the period-luminosity relation. No disagreement was found with the observational 
identification. 

The 9.70197~d$^{-1}$ dominant mode is observationally identified as the radial first 
overtone ($\ell = 0$, $n = 1$) mode but the preliminary model including low-value rotational effects gives $g3$ $\ell = 2$ identification 
for it. The latter identification is consistent with the triplet structure found around 
the dominant mode. However, this identification must be taken with caution because it
represents a minimization of the modelling, which requires 
refining rotation effects. It also modifies the other physical
parameters considered (metallicity, convective parameters, etc.) to
which the identification is likely quite sensitive. 
The radial fundamental (7.54062~d$^{-1}$), the radial second overtone 
(12.24848~d$^{-1}$,) and the $\ell = 2$, $n = 1$ (14.24347~d$^{-1}$) modes have the 
same identification from the observation and from modelling.

The low periodicities found are too low to be connected with rotation;
however, we cannot excluded that 
they are due to some rotational effects (like asymmetries). With our current
tools (perturbative methods for
the oscillation computation), those effects cannot be properly analysed (the
rotational velocity of the star
is in the limit of validity of pertubative methods) and non-peturbative
techniques would be necessary.
Interestingly, even for deformed stars, the effect of rotation is very
small on large spacings, so
the present modelling in combination on 2D equilibrium models and
non-peturbative techniques
for the oscillation is potentially a source of important information on the
mode identification and on 
the internal rotation of the star.

The unexpected regularity of a $\delta$ Scuti star that is out of the 
asymptotic regime shows that the quasi-equidistancies can be found among the 
frequencies of space data. The potential of the mode identification based only on the 
frequency regularities is highly valuable. The space missions observed a lot of stars 
and the capacity of the ground-based spectroscopy was not enough for mode identification 
on high-resolution spectra. The tentative mode identification in each star observed in 
the non-asymptotic regime would make a step toward discovering the real nature of 
the mode-selection mechanism.

\begin{acknowledgements}
MP, ZsB, JMB, and MH acknowledge the support of the ESA PECS project 4000103541/11/NL/KML. 
EP acknowledges financial support from the PRIN-INAF 2010 {\it Asteroseismology: 
looking inside the stars with space- and ground-based observations}. 
AM acknowledge the support of the projects Consolider-CSD2006-00070, ESP2007-65475-C02-02, and
AYA2010-21161-C02-02, and the Madrid regional government grant PRICIT-S2009ESP-1496. AM acknowledge the support of the Comunidad de Madrid grant AstroMadrid
(S2009/ESP-1496) and the projects AYA
2010-21161-C02-02,AYA2012-38897-C02-01, and PRICITS2009/ESP-1496.
\'AS acknowledges support from the Belgian Federal Science Policy
(project MO/33/029).
WW was supported by the Austrian Science Fonds, P22691-N16. JCS acknowledges support from the Spanish Ministry through National Research grant AYA2012-39346-C02-01. DG, EG are grateful to the user support group of AAT for all their help
and assistance for preparing and carrying out the observations. They
would like to particularly thank Rob Sharp, Fred Watson, and Quentin Parker.
Based on observations obtained with the HERMES spectrograph, which is
supported by the Fund for Scientific Research of Flanders (FWO),
Belgium, the Research Council of K.U.Leuven, Belgium, the Fonds
National Recherches Scientific (FNRS), Belgium, the Royal Observatory
of Belgium, the Observatoire de Geneve, Switzerland, and the Thuringer
Landessternwarte Tautenburg, Germany.
This research has made use of the Exo-Dat database, operated at LAM-OAMP,
Marseille, France, on behalf of the CoRoT/Exoplanet programme. 
\end{acknowledgements}
\bibliographystyle{aa}

\listofobjects

\Online

\longtab{9}{
\begin{longtable}{lrcrrrrrrrr}
\caption{The 94 frequencies of CoRoT\,102749568 obtained in white light, above the $sig$ parameter 5.0 calculated by SigSpec. For completeness, the amplitudes and phases are given in CoRoT blue, green and red colours, too. The frequencies No.~91--94 are considered to be technical ones.}\label{tabl:online}\\
\hline
\hline
 & \multicolumn{4}{c}{White light} & \multicolumn{2}{c}{Blue} & \multicolumn{2}{c}{Green} & \multicolumn{2}{c}{Red}\\
No. & \multicolumn{1}{c}{Frequency} & \multicolumn{1}{c}{Ampl.} & \multicolumn{1}{c}{Phase} &
\multicolumn{1}{c}{S/N} & \multicolumn{1}{c}{Ampl.} & \multicolumn{1}{c}{Phase} & \multicolumn{1}{c}{Ampl.}
& \multicolumn{1}{c}{Phase} & \multicolumn{1}{c}{Ampl.} & \multicolumn{1}{c}{Phase} \\
& \multicolumn{1}{c}{(d$^{-1}$)} & \multicolumn{1}{c}{(mmag)} & \multicolumn{1}{c}{($1/360\degr$)} & &
\multicolumn{1}{c}{(mmag)} & \multicolumn{1}{c}{($1/360\degr$)} & \multicolumn{1}{c}{(mmag)} &
\multicolumn{1}{c}{($1/360\degr$)} & \multicolumn{1}{c}{(mmag)} & \multicolumn{1}{c}{($1/360\degr$)}\\
\hline
\endfirsthead
\caption{Continued.} \\
\hline
 & \multicolumn{4}{c}{White light} & \multicolumn{2}{c}{Blue} & \multicolumn{2}{c}{Green} & \multicolumn{2}{c}{Red}\\
No. & \multicolumn{1}{c}{Frequency} & \multicolumn{1}{c}{Ampl.} & \multicolumn{1}{c}{Phase} &
\multicolumn{1}{c}{S/N} & \multicolumn{1}{c}{Ampl.} & \multicolumn{1}{c}{Phase} & \multicolumn{1}{c}{Ampl.}
& \multicolumn{1}{c}{Phase} & \multicolumn{1}{c}{Ampl.} & \multicolumn{1}{c}{Phase} \\
& \multicolumn{1}{c}{(d$^{-1}$)} & \multicolumn{1}{c}{(mmag)} & \multicolumn{1}{c}{($1/360\degr$)} & &
\multicolumn{1}{c}{(mmag)} & \multicolumn{1}{c}{($1/360\degr$)} & \multicolumn{1}{c}{(mmag)} &
\multicolumn{1}{c}{($1/360\degr$)} & \multicolumn{1}{c}{(mmag)} & \multicolumn{1}{c}{($1/360\degr$)}\\
\hline
\endhead
\hline
\endfoot
\hline
\endlastfoot
1 &  9.70197(1) & \multicolumn{1}{r}{10.51(3)} & 0.3759(4) & 300.28 & 14.60 & 0.3729 & 12.51 & 0.3764 & 8.80 & 0.3796\\
2 & 9.93909(2) &   4.77 & 0.3616(9) & 136.09 & 6.57 & 0.3467 & 5.60 & 0.3563 & 4.05 & 0.3716\\
3 & 11.64346(3) & 3.66 & 0.302(1) & 102.40 & 5.01 & 0.2960 & 4.28 & 0.3064 & 3.09 & 0.3240\\
4 & 8.30727(6) &   1.75 & 0.086(2) & 49.98 & 2.43 & 0.0837 & 2.11 & 0.0911 & 1.46 & 0.0880\\
5 & 9.55903(8) &   1.44 & 0.683(3) & 41.54 & 2.01 & 0.6833 & 1.75 & 0.6848 & 1.20 & 0.6836\\
6 & 11.53036(9) & 1.21 & 0.610(3) & 33.86 & 1.60 & 0.6010 & 1.49 & 0.6068 & 1.02 & 0.6143\\
7 & 12.2485(1) & 1.01 & 0.411(4) & 27.90 & 1.41 & 0.4125 & 1.19 & 0.4115 & 0.86 & 0.4127\\
8 & 10.8256(1) & 0.92 & 0.821(4) & 25.77 & 1.22 & 0.8033 & 1.10 & 0.8199 & 0.78 & 0.8149\\
9 & 10.6072(1) & 0.88 & 0.826(5) & 24.95 & 1.23 & 0.8074 & 1.05 & 0.8050 & 0.73 & 0.8220\\
10 & 8.37540(1) & 0.75 & 0.829(6) & 21.28 & 1.03 & 0.8382 & 0.90 & 0.8134 & 0.63 & 0.8282\\
11 & 14.2435(2) & 0.73 & 0.409(6) & 21.02 & 1.04 & 0.3999 & 0.83 & 0.4330 & 0.60 & 0.4284\\
12 & 16.7771(2) & 0.68 & 0.032(6) & 19.68 & 0.87 & 0.0367 & 0.76 & 0.0341 & 0.61 & 0.0497\\
13 & 17.5955(2) & 0.60 & 0.384(7) & 17.69 & 1.02 & 0.3727 & 0.70 & 0.3614 & 0.45 & 0.3859\\
14 & 9.3633(3) & 0.43 & 0.15(1) & 12.48 & 0.65 & 0.1409 & 0.54 & 0.1589 & 0.35 & 0.1562\\
15 & 9.9970(3) & 0.41 & 0.35(1) & 11.76 & 0.51 & 0.3581 & 0.53 & 0.3210 & 0.35 & 0.3557\\
16 & 9.1715(3) & 0.39 & 0.62(1) & 11.26 & 0.53 & 0.5981 & 0.48 & 0.6081 & 0.34 & 0.6374\\
17 & 14.0508(4) & 0.29 & 0.90(1) & 8.42 &  &  & 0.35 & 0.9325 & 0.29 & 0.8863\\
18 & 1.3807(3) & 0.33 & 0.88(1) & 8.27 & 0.44 & 0.8957 & 0.37 & 0.8857 & 0.29 & 0.8673\\
19 & 3.4271(4) & 0.31 & 0.78(1) & 8.15 & 0.47 & 0.7957 & 0.32 & 0.7573 & 0.25 & 0.7799\\
20 & 13.7356(4) & 0.27 & 0.53(2) & 7.73 & 0.31 & 0.5571 & 0.34 & 0.5693 & 0.24 & 0.5229\\
21 & 21.3933(4) & 0.25 & 0.19(2) & 7.63 &  &  & 0.29 & 0.2052 & 0.24 & 0.1805\\
22 & 10.5891(4) & 0.27 & 0.11(2) & 7.56 & 0.30 & 0.0819 & 0.31 & 0.1165 & 0.26 & 0.1303\\
23 & 6.5978(4) &  0.25 & 0.68(2) & 7.40 & 0.45 & 0.6804 & 0.33 & 0.6627 & 0.16 & 0.6827\\
24 & 12.5223(4) & 0.27 & 0.35(2) & 7.39 & 0.35 & 0.3348 & 0.32 & 0.3487 & 0.23 & 0.3542\\
25 & 8.7073(4) &  0.26 & 0.03(2) & 7.29 & 0.39 & 0.0251 & 0.27 & 0.0482 & 0.21 & 0.0372\\
26 & 7.5406(4) &  0.25 & 0.96(2) & 7.26 & 0.41 & 0.9493 & 0.25 & 0.9204 & 0.21 & 0.9725\\
27 & 20.1384(5) & 0.24 & 0.25(2) & 7.20 & 0.34 & 0.2581 &  &  & 0.21 & 0.2396\\
28 & 13.4248(5) & 0.23 & 0.52(2) & 6.64 & 0.37 & 0.4827 & 0.34 & 0.5253 & 0.19 & 0.5572\\
29 & 14.8295(5) & 0.23 & 0.77(2) & 6.54 & 0.35 & 0.7904 & 0.25 & 0.7867 & 0.18 & 0.7711\\
30 & 14.8163(5) & 0.23 & 0.32(2) & 6.51 &  &  & 0.26 & 0.2573 & 0.27 & 0.3456\\
31 & 10.1663(5) & 0.23 & 0.33(2) & 6.48 & 0.31 & 0.3343 & 0.28 & 0.3261 & 0.19 & 0.3493\\
32 & 1.3317(6) & 0.26 & 0.66(2) & 6.48 & 0.39 & 0.6408 & 0.33 & 0.6430 & 0.22 & 0.6510\\
33 & 14.8818(5) & 0.22 & 0.55(2) & 6.46 & 0.29 & 0.5412 & 0.22 & 0.5546 & 0.21 & 0.5321\\
34 & 10.0113(6) & 0.22 & 0.67(2) & 6.19 & 0.36 & 0.7020 & 0.26 & 0.6378 & 0.16 & 0.6392\\
35 & 23.0662(6) & 0.19 & 0.75(2) & 6.04 &  &  &  &  & 0.23 & 0.7604\\
36 & 1.4412(5) & 0.24 & 0.07(2) & 6.03 & 0.34 & 0.0438 & 0.30 & 0.0667 & 0.21 & 0.0830\\
37 & 2.6323(5) & 0.22 & 0.08(2) & 5.78 & 0.29 & 0.1225 & 0.28 & 0.0845 & 0.19 & 0.0518\\
38 & 17.4440(6) & 0.20 & 0.07(2) & 5.74 & 0.29 & 0.0638 & 0.24 & 0.0670 & 0.16 & 0.0655\\
39 & 3.9661(6) & 0.20 & 0.92(2) & 5.41 & 0.30 & 0.8790 & 0.29 & 0.8912 & 0.15 & 0.9650\\
40 & 19.2125(6) & 0.18 & 0.34(2) & 5.38 & 0.29 & 0.3964 &  &  & 0.16 & 0.3169\\
41 & 3.9804(6) & 0.20 & 0.85(2) & 5.37 &  &  &  &  & 0.19 & 0.8278\\
42 & 6.3053(6) & 0.18 & 0.30(2) & 5.33 &  &  & 0.23 & 0.3578 & 0.15 & 0.2925\\
43 & 10.7629(6) & 0.19 & 0.37(2) & 5.29 & 0.31 & 0.3681 &  &  & 0.16 & 0.4008\\
44 & 1.9396(6) & 0.21 & 0.97(2) & 5.28 & 0.29 & 0.0417 & 0.25 & 0.9921 & 0.18 & 0.9333\\
45 & 11.4072(6) & 0.19 & 0.63(2) & 5.25 & 0.34 & 0.5835 &  &  & 0.16 & 0.6748\\
46 & 11.5837(6) & 0.19 & 0.38(2) & 5.22 &  &  &  &  & 0.16 & 0.3844\\
47 & 18.7279(7) & 0.17 & 0.93(2) & 5.18 &  &  &  &  & 0.18 & 0.9589\\
48 & 1.4097(6) & 0.20 & 0.15(2) & 5.09 & 0.29 & 0.1546 & 0.23 & 0.1474 & 0.17 & 0.1600\\
49 & 14.1922(6) & 0.17 & 0.95(2) & 4.92 & 0.31 & 0.8875 & 0.21 & 0.9290 & 0.13 & 0.9910\\
50 & 5.6627(6) &  0.18 & 0.33(2) & 4.91 &  &  & 0.24 & 0.2742 & 0.17 & 0.3547\\
51 & 22.7201(7) & 0.16 & 0.58(3) & 4.88 &  &  &  &  & 0.21 & 0.5800\\
52 & 1.7036(6) & 0.19 & 0.13(2) & 4.82 &  &  &  &  & 0.19 & 0.1385\\
53 & 10.6974(7) & 0.17 & 0.74(2) & 4.72 & 0.31 & 0.7431 &  &  &  & \\
54 & 16.6673(7) & 0.16 & 0.71(3) & 4.67 &  &  &  &  & 0.17 & 0.6971\\
55 & 21.5762(7) & 0.15 & 0.13(3) & 4.63 &  &  &  &  & 0.14 & 0.1489\\
56 & 16.3344(7) & 0.16 & 0.72(3) & 4.57 &  &  &  &  & 0.15 & 0.7376\\
57 & 17.0643(7) & 0.15 & 0.56(3) & 4.39 &  &  &  &  & 0.17 & 0.5906\\
58 & 5.6105(7) & 0.16 & 0.22(3) & 4.38 &  &  &  &  & 0.15 & 0.2058\\
59 & 15.2310(7) & 0.15 & 0.80(3) & 4.31 &  &  &  &  & 0.14 & 0.7996\\
60 & 2.6950(7) & 0.16 & 0.36(3) & 4.28 &  &  &  &  & 0.15 & 0.3837\\
61 & 18.1188(8) & 0.14 & 0.29(3) & 4.26 &  &  &  &  & 0.13 & 0.2977\\
62 & 21.7796(8) & 0.14 & 0.72(3) & 4.19 &  &  &  &  & 0.15 & 0.7360\\
63 & 3.2651(7) & 0.16 & 0.99(2) & 4.18 &  &  & 0.24 & 0.9865 & 0.14 & 0.0058\\
64 & 1.325(1) & 0.16 & 0.19(3) & 4.12 &  &  &  &  &  & \\
65 & 8.8190(8) & 0.14 & 0.38(3) & 4.11 &  &  &  &  & 0.13 & 0.3804\\
66 & 14.4294(7) & 0.14 & 0.07(3) & 4.10 &  &  &  &  &  & \\
67 & 16.4689(8) & 0.14 & 0.88(3) & 4.01 &  &  &  &  &  & \\
68 & 12.7913(8) & 0.14 & 0.46(3) & 3.93 &  &  &  &  & 0.15 & 0.4437 \\
69 & 3.1978(7) &  0.15 & 0.12(3) & 3.86 &  &  &  &  &  &  \\
70 & 9.9855(8) &  0.14 & 0.17(3) & 3.84 &  &  &  &  &  &  \\
71 & 0.0468(8) &  0.15 & 0.60(3) & 3.83 &  &  &  &  &  &  \\
72 & 12.3629(8) & 0.14 & 0.58(3) & 3.80 & 0.28 & 0.5726 &  &  &  &  \\
73 & 14.5637(8) & 0.13 & 0.13(3) & 3.79 &  &  &  &  &  &  \\
74 & 9.8822(8) &  0.13 & 0.50(3) & 3.79 &  &  &  &  &  &  \\
75 & 16.8315(8) & 0.13 & 0.44(3) & 3.78 &  &  &  &  & 0.13 & 0.4704 \\
76 & 0.2381(7) &  0.15 & 0.70(3) & 3.66 &  &  &  &  &  &  \\
77 & 19.2988(9) & 0.12 & 0.82(3) & 3.65 &  &  &  &  &  &  \\
78 & 13.9927(9) & 0.13 & 0.51(3) & 3.63 &  &  & 0.40 & 0.4708 & 0.07 & 0.8553 \\
79 & 7.7238(9) &  0.13 & 0.80(3) & 3.61 &  &  &  &  & 0.14 & 0.8040 \\
80 & 9.2103(9) &  0.12 & 0.33(3) & 3.59 &  &  &  &  &  &  \\
81 & 11.4228(9) & 0.13 & 0.90(3) & 3.59 &  &  &  &  &  &  \\
82 & 15.0957(9) & 0.12 & 0.53(3) & 3.57 &  &  &  &  & 0.14 & 0.5317 \\
83 & 3.0410(8) &  0.14 & 0.90(3) & 3.56 & 0.28 & 0.7700 &  &  & 0.13 & 0.9924 \\
84 & 3.6837(8) &  0.13 & 0.17(3) & 3.48 &  &  &  &  &  &  \\
85 & 2.9902(8) &  0.13 & 0.24(3) & 3.45 &  &  &  &  & 0.15 & 0.2902 \\
86 & 2.7296(8) &  0.13 & 0.69(3) & 3.40 &  &  &  &  & 0.14 & 0.6755 \\
87 & 2.3200(8) &  0.13 & 0.22(3) & 3.37 &  &  &  &  &  &  \\
88 & 0.0655(8) &  0.13 & 0.04(3) & 3.33 &  &  &  &  &  &  \\
89 & 0.2582(8) &  0.13 & 0.84(3) & 3.29 &  &  &  &  &  &  \\
90 & 3.3002(9) &  0.12 & 0.33(3) & 3.22 &  &  &  &  &  &  \\
 &  &  &  &  &  &  &  &  &  &  \\
91 & 41.911(2) & 0.28 & 0.15(3) & 9.22 & 0.30 & 0.0553 & 0.25 & 0.0412 & 0.12 & 0.1634 \\
92 & 41.907(2) & 0.22 & 0.09(3) & 7.25 &  &  &  &  &  &  \\
93 & 55.8811(7) & 0.16 & 0.30(3) & 5.53 & 0.40 & 0.2902 & 0.41 & 0.4092 & 0.11 & 0.1492 \\
94 & 27.9493(7) & 0.15 & 0.69(3) & 4.88 & 0.40 & 0.5244 & 0.27 & 0.3918 &  &  \\

\end{longtable}
}
 
\end{document}